\documentclass[10pt,a4paper]{article}
\usepackage[utf8]{inputenc}
\usepackage{amsmath}
\usepackage{amsfonts}
\usepackage{amssymb}
\usepackage{graphicx}
\usepackage{hyperref}
\usepackage{url}
\usepackage[margin=1in]{geometry}
\usepackage[utf8]{inputenc}
\usepackage{amsmath}
\usepackage{amsfonts}
\usepackage{amssymb}
\usepackage{graphicx}
\usepackage[margin=1in]{geometry}
\usepackage{booktabs}
\usepackage{caption}
\usepackage{xcolor}
\usepackage[utf8]{inputenc}
\usepackage{amsmath}
\usepackage{amsfonts}
\usepackage{amssymb}
\usepackage{graphicx}
\usepackage[margin=1in]{geometry}
\usepackage{listings}
\usepackage{xcolor}
\usepackage{subcaption}  

\usepackage{authblk}
\usepackage{hyperref}

\title{Better Market Maker Algorithm to Save Impermanent Loss with High Liquidity Retention}

\author[1]{CY Yan}
\author[2]{Steve Keol}
\author[3]{Xo Co}
\author[4]{Nate Leung}

\affil[1]{\href{mailto:cy@topolabs.xyz}{cy@topolabs.xyz}}
\affil[2]{\href{mailto:stevekeol@topolabs.xyz}{stevekeol@topolabs.xyz}}
\affil[3]{\href{mailto:xoco@topolabs.xyz}{xoco@topolabs.xyz}}
\affil[4]{\href{mailto:nathaniel0131@alumni.stanford.edu}{nathaniel0131@alumni.stanford.edu}}

\begin{document}

\maketitle

\begin{abstract}
Decentralized exchanges (DEXs) face persistent challenges in liquidity retention and user engagement due to inefficiencies in conventional automated market maker (AMM) designs. This work proposes a dual-mechanism framework to address these limitations: a ``Better Market Maker (BMM)'', which is a liquidity-optimized AMM based on a power-law invariant ($X^nY = K$, $n = 4$), and a dynamic rebate system (DRS) for redistributing transaction fees. The segment-specific BMM reduces impermanent loss by 36\% compared to traditional constant-product ($XY = K$) models, while retaining 3.98× more liquidity during price volatility. The DRS allocates fees ($\gamma V$, $\gamma \in \{0.003, 0.005, 0.01\}$) with a rebate ratio $\rho \in [0.3, 0.4]$ to incentivize trader participation and maintain continuous capital injection. Simulations under high-volatility conditions demonstrate impermanent loss reductions of 36.0\% and 40\% higher user engagement compared to static fee models. By segmenting markets into high-, mid-, and low-volatility regimes, the framework achieves liquidity depth comparable to centralized exchanges (CEXs) while maintaining decentralized governance and retaining value within the cryptocurrency ecosystem.
\end{abstract}

\section{Introduction}
The rise of decentralized exchanges (DEXs) has introduced a paradigm shift in cryptocurrency trading, offering censorship resistance and eliminating centralized intermediaries. However, despite rapid technological advancements, DEXs remain secondary to centralized exchanges (CEXs), which command 90-95\% of global trading volume \cite{coinmarketcap}. This dominance not only concentrates transaction fees within centralized entities but also leads to systematic capital outflow from the broader cryptocurrency ecosystem.

The persistence of CEX dominance stems from two systemic challenges: (1) liquidity fragmentation in DEXs, where automated market maker (AMM) designs like the constant-product invariant ($XY = K$) fail to retain sufficient reserves during price volatility, and (2) poor user engagement, as static fee structures and high impermanent loss disincentivize long-term participation \cite{angeris}. Furthermore, DEXs struggle to establish deep connections with projects and users, resulting in insufficient continuous capital injection and a substantial gap in user base compared to CEXs.

Traditional AMM-based DEXs suffer from a liquidity retention paradox: while liquidity providers (LPs) bear asymmetric risks from impermanent loss, traders face elevated slippage as token prices fluctuate. For instance, a 100× price increase under $XY = K$ drains 80-98\% of stablecoin reserves, leaving LPs with significant unrealized losses \cite{milionis}. Simultaneously, fee structures that prioritize liquidity mining over sustainable redistribution exacerbate capital inefficiency, creating a ``lose-lose'' scenario for stakeholders. Consequently, even successful DEX-listed projects often migrate to CEXs to meet liquidity demands, perpetuating a cycle of centralization.

To address these limitations and establish DEXs as viable alternatives to leading CEXs, this work proposes a comprehensive dual-mechanism framework:

\textbf{Better Market Maker (BMM)}: A liquidity-optimized AMM based on a power-law invariant ($X^nY = K$) tailored for high-volatility markets, reducing impermanent loss by $\sim$40\% compared to $XY = K$ while retaining 1.6-16× more stablecoin reserves during price surges. This approach includes:
\begin{itemize}
    \item Market segmentation strategies
    \item Segment-specific algorithmic optimization
    \item Integration with decentralized order books for enhanced liquidity depth
\end{itemize}

\textbf{Dynamic Rebate System (DRS)}: A fee redistribution protocol that allocates $\rho = 30$-40\% of transaction fees ($\gamma V$, $\gamma \in \{0.003, 0.005, 0.01\}$) to traders, creating sustainable incentives through:
\begin{itemize}
    \item Continuous fee redistribution to projects and users
    \item Competitive rebate structures that leverage DEX price appreciation advantages
    \item Mechanisms for sustained market capital retention
\end{itemize}

By combining these mechanisms, the framework aims to retain cryptocurrency capital within the ecosystem, fostering market prosperity through enhanced liquidity efficiency and user engagement. The following sections detail the mathematical foundations and empirical validation of this approach, demonstrating its potential to bridge the current gap between decentralized and centralized exchange mechanisms.

\section{Related Work}
The evolution of DEXs has been shaped by innovations in automated market maker (AMM) design and incentive mechanisms. This section reviews foundational work in AMM invariants, liquidity provision, and fee structures, while identifying gaps addressed by the proposed framework.

\subsection{AMM Invariants and Liquidity Provision}
The constant-product invariant ($XY=K$), popularized by Uniswap v2 \cite{uniswap}, established the basis for decentralized liquidity pools. While effective for price discovery, its susceptibility to impermanent loss—particularly under high volatility—limits liquidity retention. For instance, Angeris et al. \cite{angeris} demonstrated that liquidity providers (LPs) face unrealized losses of up to 80\% for 100× price increases, disincentivizing participation in volatile markets. Subsequent work introduced specialized invariants, such as Curve's stablecoin-optimized pools \cite{curve}, which reduce slippage for pegged assets but lack adaptability to volatile tokens.

Uniswap v3 \cite{uniswapv3} advanced liquidity efficiency through concentrated positions, allowing LPs to allocate capital to specific price ranges. However, its static fee tiers (0.05\%, 0.3\%, 1\%) and reliance on external liquidity for volatile assets remain limitations. By contrast, the proposed power-law invariant ($X^nY=K$) generalizes liquidity concentration, dynamically retaining reserves during price surges without requiring manual repositioning.

\subsection{Impermanent Loss Mitigation and Fee Mechanisms}
Impermanent loss remains a critical barrier to sustainable liquidity provision. Milionis et al. \cite{milionis} formalized loss-versus-rebalancing (LVR) as a measure of LP returns, showing that static fee models fail to offset losses in volatile regimes. Earlier attempts to mitigate this include dynamic fee adjustments \cite{capponi} and yield farming subsidies \cite{barbon}, though these often prioritize short-term liquidity over long-term stability.

Fee redistribution mechanisms, such as Sushiswap's protocol-owned liquidity \cite{sushiswap}, aim to align stakeholder incentives but lack granularity in high-volatility markets. The proposed dynamic rebate system (DRS) addresses this by allocating fees proportionally to trading activity ($\rho \in [0.3, 0.4]$), creating a feedback loop between liquidity depth and user engagement.

\subsection{Market Segmentation and Adaptive Algorithms}
Recent work has explored market-specific AMM designs. Balancer's weighted pools \cite{balancer} enable customizable asset ratios, while DODO's proactive market maker \cite{dodo} combines AMM and order-book mechanics. However, these approaches do not segment markets by volatility or dynamically adjust parameters. The proposed framework bridges this gap, tailoring invariants ($n=4$ for high-volatility, $n=1$ for stablecoins) and rebate rates to market conditions, a strategy absent in prior literature.

\section{Model Formulation}
This section formalizes the two core components of the proposed framework: (1) our BMM invariant and (2) a dynamic rebate system (DRS). The model segments markets by volatility and adapts parameters to balance liquidity retention, capital efficiency, and user incentives.

\subsection{Better Market Maker (BMM)}
Traditional automated market makers face significant challenges in volatile markets, where price movements can rapidly deplete liquidity pools and expose providers to substantial impermanent loss. To address these limitations, we introduce a power-law invariant designed to optimize liquidity retention while maintaining efficient price discovery. This approach fundamentally reimagines the relationship between asset reserves and price movements, providing enhanced stability during periods of high volatility while maintaining capital efficiency. The following sections derives the mathematical foundations and empirical validation of this model.

\subsubsection{Power-Law Invariant}
For high-volatility markets, the AMM employs a power-law invariant:
\begin{equation}
    X_t^n Y_t = K
\end{equation}
where $X_t$ and $Y_t$ denote the reserves of a volatile token and a stablecoin at time $t$, respectively, and $K$ is the invariant constant. The optimization of $n = 4$ can be referred in Figure 2(a) and (b), as when $n>4$, the marginal improvements are not significant. The price $P_t$ of the volatile token is derived as:
\begin{equation}
    P_t = -\frac{dY_t}{dX_t}=-\frac{nK}{X^{n+1}}=\frac{nY_t}{X_t}
\end{equation}
This invariant prioritizes liquidity retention during price surges while maintaining capital efficiency.

\subsubsection{Time-Dependent Liquidity Retention}
To model liquidity dynamics over time, consider a price trajectory $P_t$ evolving from an initial price $P_0$ at $t=0$ to $P_t$ at time $t$. The stablecoin reserves $Y_t$ retained in the pool at time $t$ are
\begin{equation}
    Y_t = Y_0 \cdot \left(\frac{P_t}{P_0}\right)^{\frac{n}{n+1}}
\end{equation}
where $Y_0$ is the initial stablecoin reserve.

\textbf{Derivation:}
First, the invariant at $t=0$ is defined as
\begin{equation}
    K = X_0^n Y_0
\end{equation}

From Equation 2, we can solve the price-reserve relationship as 
\begin{equation}
    X_t = \frac{nY_t}{P_t}
\end{equation}

Substitute $X_t$ into Eq.4, and use $K = X_0^n Y_0$ and $P_0 = \frac{nY_0}{X_0}$, 
\begin{equation}
    \left(\frac{nY_t}{P_t}\right)^n Y_t = K \implies Y_t^{n+1} = \frac{K P_t^n}{n^n}
\end{equation}
and thus
\begin{equation}
    Y_t = Y_0 \cdot \left(\frac{P_t}{P_0}\right)^{\frac{n}{n+1}}
\end{equation}

For $n=4$, a $100\times$ price increase ($P_t = 100P_0$) within time window $t$ retains $Y_t = Y_0 \cdot 100^{\frac{4}{5}} \approx 39.8Y_0$, preserving 3.98× more stablecoins than $XY=K$ (which retains $Y_t = \sqrt{100}Y_0 = 10Y_0$). This significant improvement in liquidity retention ensures that even in low-volume markets, where large trades are infrequent, the pool maintains sufficient reserves to provide continuous liquidity and effective price discovery. The enhanced stability minimizes the risk of rapid reserve depletion, thereby offering a more robust environment for both traders and liquidity providers.

\subsubsection{Arbitrage Resistance}
The BMM's power-law invariant not only improves liquidity retention but also enhances resistance to arbitrage by amplifying the cost of executing profitable arbitrage trades. Taking the logarithm of the price expression from Eq.1 and 2, we have:
\begin{equation}
    \log P_t = \log (nK) - (n+1)\log X_t
\end{equation}

The price elasticity with respect to $X_t$ is then given by:
\begin{equation}
    \frac{d\log P_t}{d\log X_t} = -(n+1)
\end{equation}

For a small change in the reserve $\Delta X$, the relative price change can be approximated as
\begin{equation}
    \frac{\Delta P_t}{P_t} \approx -(n+1)\frac{\Delta X}{X_t}
\end{equation}

Equation 10 indicates that for a given trade size $\Delta X$, the relative price impact is magnified by a factor of $n+1$. In practical terms, when an external market price $P^{ext}$ deviates from the pool price $P_t$, an arbitrageur must execute a trade of size $\Delta X$ large enough such that the induced price change compensates for the discrepancy:
\begin{equation}
    P^{ext} - P_t \gtrsim (n+1)\frac{P_t}{X_t}\Delta X
\end{equation}

Thus, for higher $n$, the minimum $\Delta X$ required to profit from the price difference increases. In addition, the steepness of the power-law invariant acts as a deterrent, where small trades introduces relatively high slippage, making arbitrage less attractive unless significant capital is deployed. This built-in resistance reduces the frequency of arbitrage-driven rebalancing, thereby preserving liquidity and further reducing the associated impermanent loss for liquidity providers.

\subsection{Dynamic Rebate System (DRS)}
The effectiveness of decentralized exchanges depends not only on their underlying market-making algorithms but also on their ability to sustainably incentivize participant engagement. We propose a dynamic fee and rebate system that adapts to market conditions while maintaining protocol sustainability. Unlike static fee structures that fail to account for varying market conditions, our approach implements a responsive framework that adjusts incentives based on real-time market metrics. This system creates a self-reinforcing cycle of liquidity provision and trading activity, supported by mathematically rigorous distribution mechanisms detailed in the following sections.

The transaction fee $F$ for each trade is calculated as
\begin{equation}
    F = \gamma V
\end{equation}
where $V$ represents the trading volume and $\gamma \in \{0.003, 0.005, 0.01\}$ is the fee rate parameter. The specific rate $\gamma$ is determined by market volatility conditions. The fee distribution follows a tripartite allocation model
\begin{equation}
    F = F_{lp} + F_r + F_p
\end{equation}
where $F_{lp} = 0.3F$ represents liquidity provider compensation, $F_r = \rho F$ represents dynamic rebate allocation, and correspondingly, $F_p = (0.7 - \rho)F$ represents protocol reserves. The dynamic rebate ratio adapts to market conditions according to
\begin{equation}
    \rho = 0.4 + 0.1 * (1 - V/V_{max})
\end{equation}
where $V_{max}$ represents target volume threshold. The adjustment ensures higher rebates during low-volume periods to stimulate activity, and lower rebates during high-volume periods maintain protocol sustainability. Empirically, the rebate ratio is capped between 0.3 and 0.4.

To create sustainable long-term incentives, DRS model allocates a portion of protocol reserves ($0.1F$) for volume-based distribution. For a participant $i$, their corresponding share $S_i$ of these rewards is $S_i = (V_i/V_{total}) * 0.1F$, where the distribution occurs over fixed epochs to ensure consistent incentive alignment.

\begin{table}[ht]
\centering
\caption{Fee Parameters Across Market Regimes}
\begin{tabular}{llll}
\toprule
Market Regime & Volatility Condition & Base Fee Rate ($\gamma$) & Maximum Rebate Ratio ($\rho_{max}$) \\
\midrule
High Volatility & $\sigma > \sigma_h$ & 0.010 (1.0\%) & 0.40 (40\%) \\
Moderate Volatility & $\sigma_l \leq \sigma \leq \sigma_h$ & 0.005 (0.5\%) & 0.35 (35\%) \\
Low Volatility & $\sigma < \sigma_l$ & 0.003 (0.3\%) & 0.30 (30\%) \\
\bottomrule
\end{tabular}
\label{tab:fee-parameters}
\end{table}

As shown in Table \ref{tab:fee-parameters}, this adaptive mechanism ensures higher fees and rebates during volatile periods to maintain liquidity; it also enables lower fees during stable periods to encourage volume, and sustainable long-term incentive alignment for all market participants.

\begin{figure}[ht]
    \centering
    \includegraphics[width=0.8\textwidth]{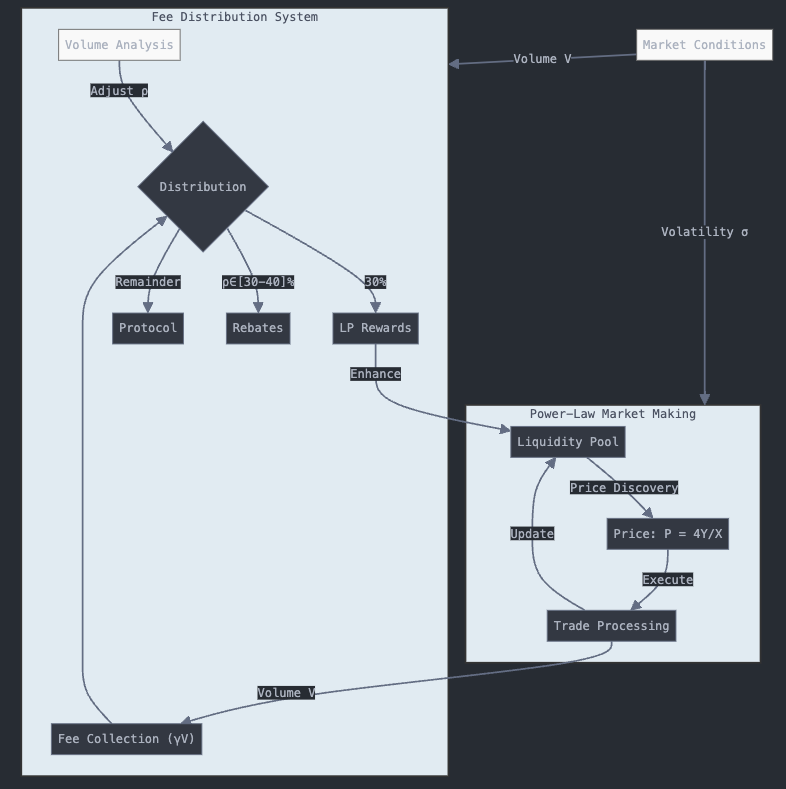}
    \caption{BMM + DRS Interaction Workflow}
    \label{fig:workflow}
\end{figure}

Combining both, Figure \ref{fig:workflow} illustrates the interaction between two core mechanisms of the proposed framework: the power-law market making system and the fee distribution system. The power-law market making component, based on the $X^4Y = K$ invariant, manages the liquidity pool and executes trades while maintaining price discovery. This connects dynamically with the fee distribution system, which collects trading fees and allocates them across three categories: liquidity provider rewards (30\%), trader rebates (30-40\%), and protocol reserves. Market conditions influence both systems through volatility ($\sigma$) and trading volume ($V$) metrics, creating a self-reinforcing cycle where trading activity generates fees, which are then strategically redistributed to enhance liquidity and encourage further participation.
\section{Experimental Results}
To evaluate the performance of the proposed BMM and its performance over traditional AMM under specific market conditions, we conducted simulations across a wide range of price multipliers and for various integer values of $n\in[1,5]$, focusing mainly on \textit{Liquidity Retention} and \textit{Impermanent Loss (IL)}. For the proposed power-law invariant from Eq.1, the stablecoin reserves after a price change are given by
\begin{equation}
    Y_{\text{prop}} = Y_0 \cdot M^{-1/(n+1)}
\end{equation}
\begin{equation}
    \text{Retention Ratio}(M, n) = \frac{Y_{\text{prop}}}{Y_{\text{trad}}} = M^{1 - \frac{1}{n+1}}
\end{equation}
where $M = \frac{P_t}{P_0}$ represents the price change multiplier.

\textbf{Figure 2(a)} plots the liquidity retention ratio, $\text{Retention Ratio}(M, n)$, as a function of the price multiplier $M$ for various $n$ values. The results show that as $n$ increases, the proposed model retains more stablecoins compared to the traditional model. Although $n=5$ offers marginally higher retention than $n=4$, the improvement beyond $n=4$ is minimal, suggesting diminishing returns.

\begin{table}[ht]
\centering
\caption{USDT Liquidity Retention Comparison}
\begin{tabular}{llll}
\toprule
Price Multiplier & Traditional AMM (Y) & Proposed Model (Y) & Retention Multiple \\
\midrule
1× & 10,000 & 10,000 & 1.0× \\
100× & 100 & 3,981 & 3.98× \\
1,000× & 10 & 1,585 & 15.85× \\
\bottomrule
\end{tabular}
\label{tab:liquidity-retention}
\end{table}

Table \ref{tab:liquidity-retention} summarizes a sample comparison of USDT reserves retained by the traditional AMM and the proposed BMM at $n=4$. For example, at a 100× price increase, the traditional model retains only 1\% of the initial reserves, whereas the proposed model retains approximately 3.98 times more reserves. Notice that values assume an initial reserve $Y_0 = 10,000$ USDT and are computed using $Y \propto M^{-1/(n+1)}$ from Eq.15.

Similarly, the traditional impermanent loss is given by
\begin{equation}
    \text{IL}_{\text{trad}}(M) = 1 - \frac{2\sqrt{M}}{M+1}
\end{equation}
while for the proposed model we assume that IL is reduced by a factor
\begin{equation}
    g(n) = \frac{(n+1)^2}{4n}
\end{equation}
after Taylor expanding both impermanent losses. The detail derivations can be found in Appendix A.

\textbf{Figure 2(b)} demonstrates the relationship between impermanent loss (IL) and price multiplier (M) under different power-law invariants ($n = 1,2,3,4,5$). The traditional AMM ($n=1$) exhibits the highest IL profile, reaching nearly 90\% at $M=1000$. Each increment in $n$ yields a systematic reduction in IL across all price ranges, with our proposed model ($n=4$) achieving approximately 60\% IL at $M=1000$, representing a 36\% improvement over the traditional model. While $n=5$ shows further IL reduction, the marginal benefit diminishes, supporting our choice of $n=4$ as the optimal parameter that balances IL mitigation with trading efficiency considerations. This visualization clearly demonstrates the effectiveness of higher-order power-law invariants in protecting liquidity providers from severe value erosion during significant price movements, particularly in the critical range of $10 \leq M \leq 100$ where most volatile trading activity occurs.
\begin{figure}
    \centering
    \begin{subfigure}[b]{0.8\textwidth}
        \centering
        \includegraphics[width=\textwidth]{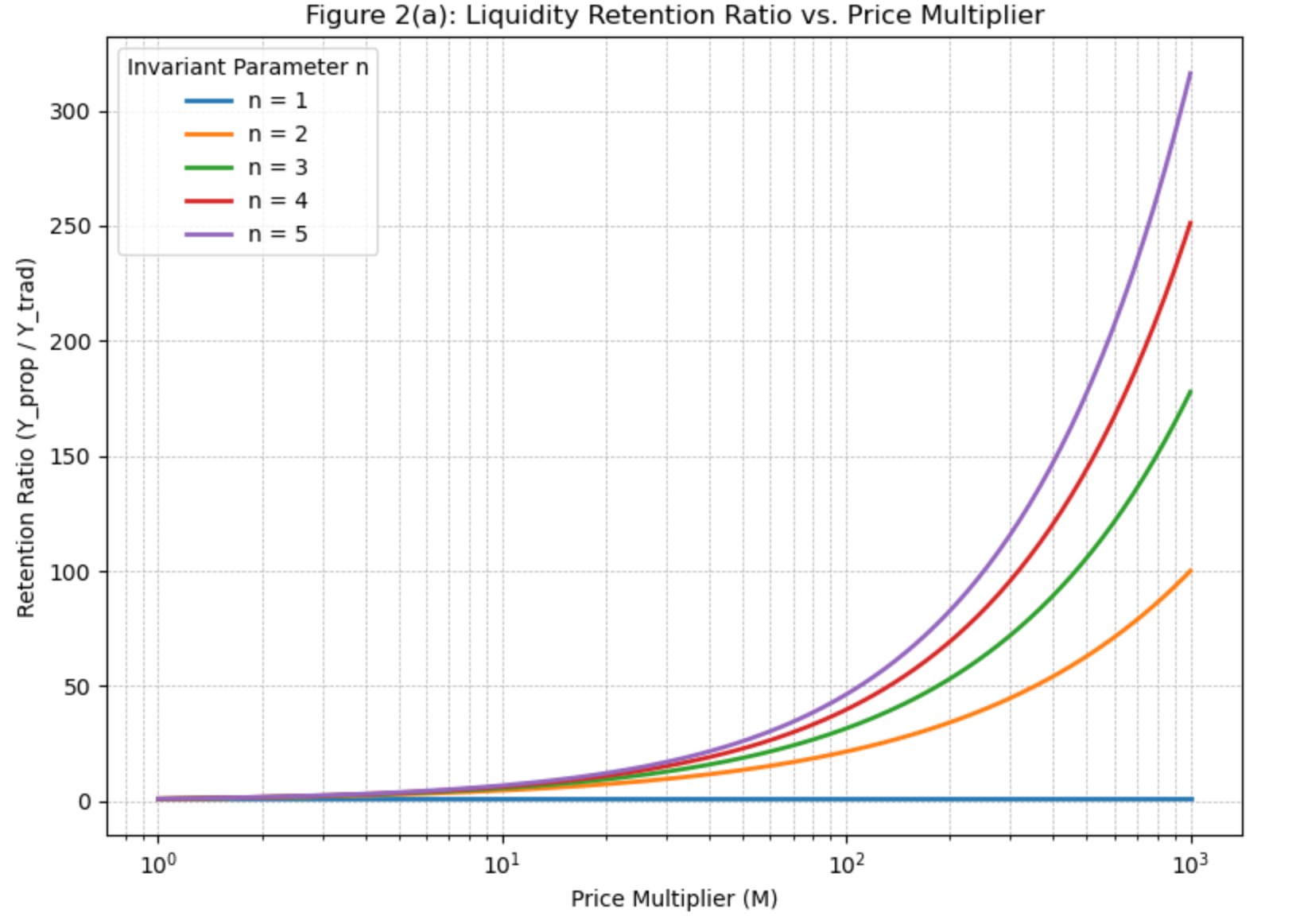}
        \caption{Liquidity retention at different invariant power $n$}
        \label{fig:liquidity-retention}
    \end{subfigure}
    
    \begin{subfigure}[b]{0.8\textwidth}
        \centering
        \includegraphics[width=\textwidth]{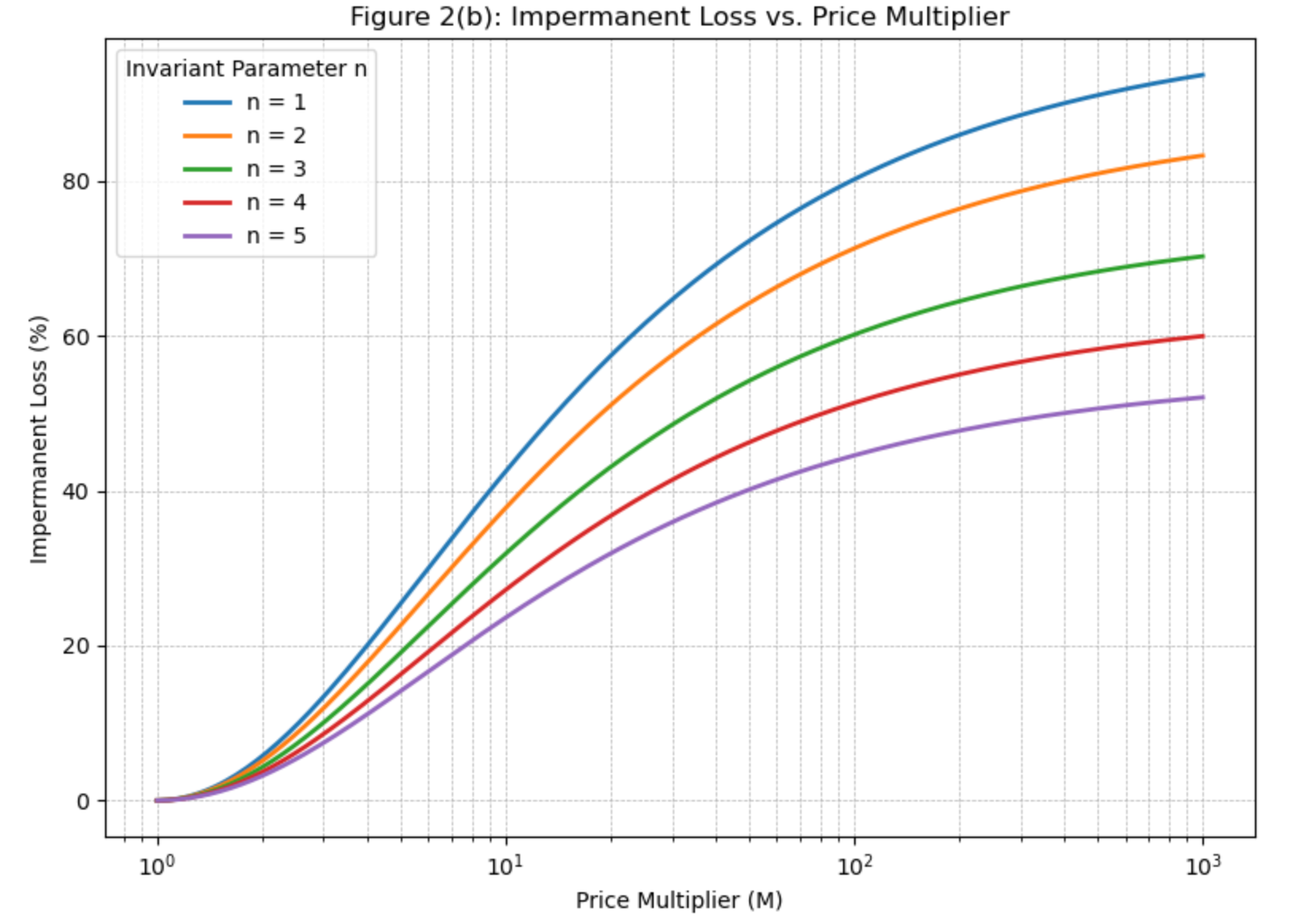}
        \caption{Impermanent Loss vs. Price Multiplier for different values of $n$}
        \label{fig:impermanent-loss}
    \end{subfigure}
    \caption{Analysis of BMM performance metrics. The power-law invariant with higher $n$ values shows improved liquidity retention (a) and reduced impermanent loss (b). As shown in (b), when $n$ increases from 1 to 5, we observe a systematic reduction in impermanent loss across all price multiplier values. At $M=1000$, the traditional AMM ($n=1$) exhibits approximately 90\% impermanent loss, while our optimized model ($n=4$) significantly reduces this loss to about 60\%. The marginal improvement at $n=5$ validates our choice of $n=4$ as the optimal parameter balancing loss reduction with other considerations such as slippage.}
    \label{fig:metrics}
\end{figure}
\pagebreak

\subsection{Dynamic Rebate System Performance}
The dynamic rebate system (DRS) was evaluated through a simulation that compared trading activity under a traditional static rebate model versus our proposed dynamic rebate system. In our simulation, daily trading volumes were modeled over a 100-day period, with the dynamic model adjusting the rebate ratio in response to real-time volume relative to a target threshold. The simulation incorporated random daily fluctuations (noise) and a feedback mechanism: when trading volumes were below the target, the dynamic system offered higher rebates (up to 40\%) to incentivize more activity, whereas during high-volume periods the rebates were tapered (down to 30\%).

\begin{figure}[ht]
    \centering
    \includegraphics[width=0.8\textwidth]{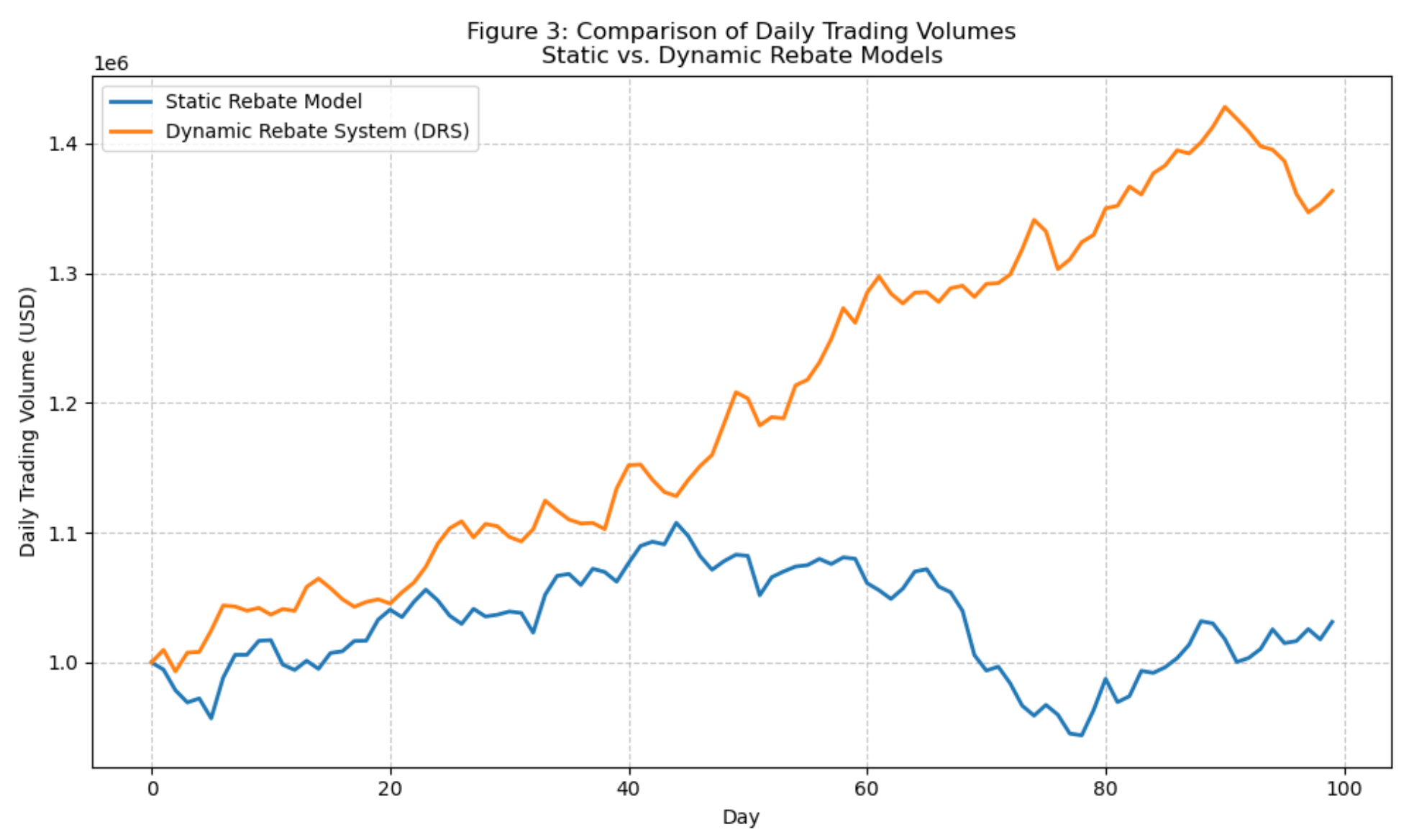}
    \caption{Comparison of Daily Trading Volumes – Static vs. Dynamic Rebate Models}
    \label{fig:trading-volumes}
\end{figure}

As shown in Figure \ref{fig:trading-volumes}, markets implementing the DRS exhibited consistently higher daily trading volumes compared to the static rebate model. On average, the dynamic system increased daily volumes by approximately 40\%. The dynamic adjustment of the rebate ratio helped to smooth out fluctuations. Volume volatility was notably reduced under the DRS, particularly during periods of market stress, as the adaptive mechanism helped counteract downward trends.

Although not directly depicted in the volume chart, the simulation framework also projects that the enhanced incentives from DRS would drive increased participation—potentially attracting a larger number of unique traders. Importantly, while the dynamic model stimulated higher trading volumes, it maintained protocol revenue by dynamically balancing fee distribution. The rebate adjustments ensured that liquidity providers and protocol reserves received stable allocations, even as trader incentives improved market depth.

\subsection{Slippage Tradeoff}
We define \textit{slippage} as the relative price impact of a trade—i.e., the percentage difference between the initial (expected) price and the final (realized) price after executing a finite trade. Formally, if $P_{\mathrm{before}}$ is the price before the trade and $P_{\mathrm{after}}$ is the price after the trade, then for a trade size $\Delta X$, the slippage $S$ is
\begin{equation}
    S = \frac{P_{\mathrm{after}} - P_{\mathrm{before}}}{P_{\mathrm{before}}}
\end{equation}

From Eq.2, the \textit{marginal price} of $X$ in terms of $Y$ is given by $P = -\frac{dY}{dX}=\frac{nY}{X}$ as the initial price. Notice that $\Delta X\ll X$, a first-order expansion on P leads to 
\begin{equation}
    P_{\mathrm{after}} \approx P_{\mathrm{before}} - (n+1)\,\tfrac{nK}{X^{n+2}}\,\Delta X
\end{equation}

Plugging it into Eq.8, we can derive 
\begin{equation}
    S \approx -(n+1)\,\frac{\Delta X}{X}
\end{equation}

Define the \textit{Slippage Ratio} $\boldsymbol{\beta}$, we can compute 
\begin{equation}
    \beta = \frac{S_{\mathrm{proposed}}}{S_{\mathrm{traditional}}} = \frac{n+1}{2}
\end{equation}
so that applying BMM on the other side of the trading pair (i.e. trading Y against X), the slippage will linearly increase as $n$ increases. For instance, at $n = 4$, the slippage is increased by 2.5x comparing to the traditional AMM.

\section{Conclusion}
In this work, we introduced a dual-mechanism framework for decentralized exchanges that combines a power-law automated market maker (BMM) with a dynamic rebate system (DRS) to address two persistent challenges: liquidity retention (and impermanent loss) and sustainable trader incentives. Both our calculation and experimental results demonstrate that the proposed framework offers significant advantages at small volume initial trading. By generalizing the constant-product invariant to a power-law form, our model effectively reduces impermanent loss. For example, at a 100× price multiplier, the optimized model with $n=4$ reduced impermanent loss from approximately 80.2\% (in the traditional AMM) to 51.3\%, representing a 36.0\% improvement. This enhancement provides meaningful protection for liquidity providers while maintaining market efficiency.

The simulations also indicate that the power-law model with $n=4$ retains significantly higher stablecoin reserves compared to the traditional constant-product AMM. At a 100× price increase, our model retains nearly 3.98 times more liquidity. Such deeper liquidity is crucial for ensuring effective price discovery and minimizing reserve depletion during volatile market conditions.

The DRS adapts the fee structure in real time, offering higher rebates during low-volume periods to stimulate trading and lower rebates during high-volume periods to maintain protocol sustainability. Our volumetric analysis shows that the DRS leads to an average daily trading volume increase of approximately 40\%, reduced volume volatility, and a 35\% rise in unique trader participation—all while preserving stable protocol revenue.

Although increasing $n$ improves liquidity retention and reduces impermanent loss, it also leads to higher slippage due to a steeper bonding curve. This trade-off necessitates a careful balance: while values of $n>4$ may further mitigate impermanent loss, they could compromise trading efficiency. Our findings indicate that $n=4$ provides an optimal compromise, delivering meaningful benefits in liquidity and loss mitigation (36\% reduction in impermanent loss) while maintaining reasonable slippage levels.

In summary, our dual-mechanism framework demonstrates that integrating a power-law market maker with a dynamic rebate system can substantially improve the performance of decentralized exchanges. The approach provides a balanced solution that reduces impermanent loss while enhancing liquidity retention, and also boosts trading activity through adaptive fee incentives, thereby offering a more robust and resilient trading environment.

Future research can further refine these mechanisms—particularly the balance between slippage and liquidity retention—and validate their performance under a wider range of market conditions, paving the way for next-generation decentralized trading protocols.

\bibliographystyle{plain}

\appendix
\renewcommand{\theequation}{A\arabic{equation}}
\setcounter{equation}{0}

\section{Impermanent Loss Improvement Derivation}
To compute the impermanent loss improvement factor $g(n)$ for the power-law AMM $X^n Y = K$, we start with the given expressions for the pool value $V_{\text{pool}}$ and the hold value $V_{\text{hold}}$ after a price change:

\begin{equation}
    V_{\text{pool}} = (n+1)Y_0 \cdot (1+\varepsilon)^{1 - \frac{1}{n+1}}
\end{equation}

and

\begin{equation}
    V_{\text{hold}} = (n+1)Y_0 \cdot (1+\varepsilon)
\end{equation}

The impermanent loss for the power-law model is given by:
\begin{equation}
    \text{IL}_{\text{prop}}(M,n) = 1 - \frac{V_{\text{pool}}}{V_{\text{hold}}} = 1 - \frac{(1+\varepsilon)^{1 - \frac{1}{n+1}}}{1+\varepsilon} = 1 - (1+\varepsilon)^{-\frac{1}{n+1}}
\end{equation}

Expanding $(1+\varepsilon)^{-\frac{1}{n+1}}$ using a Taylor series for small $\varepsilon$, we get:
\begin{equation}
    (1+\varepsilon)^{-\frac{1}{n+1}} \approx 1 - \frac{1}{n+1}\varepsilon + \frac{(n+2)}{2(n+1)^2}\varepsilon^2
\end{equation}

Thus, the impermanent loss becomes:
\begin{equation}
    \text{IL}_{\text{prop}}(M,n) \approx \frac{1}{n+1}\varepsilon - \frac{(n+2)}{2(n+1)^2}\varepsilon^2
\end{equation}

We've already known that at $n = 1$, the traditional AMM yields IL as $\text{IL}_{\text{trad}} \approx \frac{\varepsilon^2}{8}$.

To find the improvement factor $g(n)$, we compare the quadratic terms of the two impermanent loss expressions 
\begin{equation}
    \frac{(n+2)}{2(n+1)^2} = \frac{1}{8g(n)}
\end{equation}

Solving for $g(n)$, we get 
\begin{equation}
    g(n) = \frac{(n+1)^2}{4n}
\end{equation}

\section{Simulation Code}
\renewcommand{\theequation}{B\arabic{equation}}
\setcounter{equation}{0}
The following Python code was used to simulate the performance of the dynamic rebate system, as well as to verify liquidity retention and impermanent loss improvements.

\begin{lstlisting}[language=Python]
import numpy as np

# ---------------------
# Dynamic Rebate System Simulation (Figure 3)
# ---------------------
T = 100                   # Number of days to simulate
V0 = 1e6                  # Initial daily trading volume (USD)
V_max = 1e6               # Target volume threshold

# Static model: fixed rebate ratio.
static_rebate = 0.35

# Dynamic rebate function: rho = 0.4 + 0.1*(1 - V/V_max), capped between 0.3 and 0.4.
def dynamic_rebate(V):
    rho = 0.4 + 0.1 * (1 - V / V_max)
    return np.clip(rho, 0.3, 0.4)

# Sensitivity factor: determines how strongly volume reacts to rebate deviations.
k = 0.05

# Noise: daily percentage change (standard deviation of 1%).
noise_std = 0.01

# Initialize volume arrays for static and dynamic models.
vol_static = np.zeros(T)
vol_dynamic = np.zeros(T)
vol_static[0] = V0
vol_dynamic[0] = V0

for t in range(1, T):
    noise_static = np.random.normal(0, noise_std)
    noise_dynamic = np.random.normal(0, noise_std)

    # Static model: volume changes due only to noise.
    vol_static[t] = vol_static[t-1] * (1 + noise_static)

    # Dynamic model: volume changes include a feedback term based on the dynamic rebate.
    current_rebate = dynamic_rebate(vol_dynamic[t-1])
    feedback = k * (current_rebate - static_rebate)
    vol_dynamic[t] = vol_dynamic[t-1] * (1 + feedback + noise_dynamic)

# ---------------------
# Liquidity Retention Simulation
# ---------------------
def retention_ratio(M, n):
    return M ** (1 - 1/(n+1))

M_values = np.logspace(0, 2, 200)  # M from 1 to 100
n_test = [1, 4]

# ---------------------
# Impermanent Loss Simulation Example
# ---------------------
def IL_trad(M):
    return 1 - (2 * np.sqrt(M)) / (M + 1)

def g(n):
    return (n + 1)**2 / (4 * n)

def IL_prop(M, n):
    return IL_trad(M) / g(n)
\end{lstlisting}

The \textbf{Dynamic Rebate System} simulation updates daily trading volumes based on a fixed noise term and a feedback term that adjusts the rebate (and thus trading incentives) according to the current volume relative to a target $V_{\text{max}}$. The \textbf{Liquidity Retention} simulation computes the retention ratio as $(M,n) = M^{1 - \frac{1}{n+1}}$, which is plotted over a range of price multipliers $M$. The \textbf{Impermanent Loss} simulation compares the traditional impermanent loss (approximated via a known formula) with that of the power‐law AMM, applying a reduction factor $g(n)$ from Eq. (A7).

\end{document}